\documentclass[a4paper,11pt]{article}
\pdfoutput=1 % if your are submitting a pdflatex (i.e. if you have
\usepackage{jheppub} % for details on the use of the package, please
\usepackage{color}%to be able to use colors
\usepackage{amsmath}%Recomended for mathematical documents (use of: eqref,...)
\usepackage{amsfonts}
\usepackage{graphicx}
\usepackage{amssymb}
\usepackage{float}
\usepackage{comment}
\usepackage{appendix}
\usepackage[dvipsnames]{xcolor}
\definecolor{refs}{RGB}{245,156,74}

\hypersetup{
  colorlinks = true,
  linkcolor = red,
  urlcolor  = magenta,
  citecolor = blue,
  anchorcolor = blue
}

\newcommand{\dd}{{\rm d}}

\newcommand{\Lag}{\mathcal{L}}

\newcommand{\mK}{\mathcal{K}}

\newcommand{\mQ}{\mathcal{Q}}

\newcommand{\Phit}{\tilde{\Phi}}

\newcommand{\kt}{\tilde{k}}

\newcommand{\mP}{m_{\rm P}}
\newcommand{\mA}{m_{\text{A}}}

\newcommand{\Ap}{A^+}
\newcommand{\Am}{A^-}
\newcommand{\ap}{a^+}
\newcommand{\am}{a^-}

\newcommand{\Bp}{B^+}
\newcommand{\Bm}{B^-}
\newcommand{\bp}{b^+}
\newcommand{\bm}{b^-}

\newcommand{\rs}{r_{\text s}}
\newcommand{\cBI}{c_{\text{BI}}}
\newcommand{\be}{\begin{equation}}
\newcommand{\ee}{\end{equation}}
\newcommand{\bea}{\begin{eqnarray}}
\newcommand{\eea}{\end{eqnarray}}

\renewcommand{\bf}[1]{{\textbf{#1}}}
\renewcommand{\tt}[1]{\textit{#1}}

\begin{document}

\title{Finite size effects in DBI and Born-Infeld  for screened spherically symmetric objects}

\author[a,b,c]{Jose Beltr\'an Jim\'enez,}
\author[b,d]{Dario Bettoni}
\author[e]{Philippe Brax,}
\author[f]{Bert Janssen,}
\author[f]{Pablo Sampedro}

\affiliation[a]{Departamento de F\'isica Fundamental, Universidad de Salamanca, E-37008 Salamanca, Spain.}
\affiliation[b]{Instituto Universitario de Física Fundamental y Matemátivas (IUFFyM), Universidad de Salamanca, E-37008 Salamanca, Spain.}
\affiliation[c]{Institute of Theoretical Astrophysics, University of Oslo, N-0315 Oslo, Norway.}
\affiliation[d]{Departamento de Matemáticas, Universidad de León,
Escuela de Ingenierías Industrial, Informática y Aeroespacial
Campus de Vegazana, s/n
24071 León, Spain.}
\affiliation[e]{Institut de Physique Th\'eorique, Universit\'e Paris-Saclay, CEA, CNRS, F-91191 Gif-sur-Yvette Cedex, France.}

\affiliation[f]{Departamento de F\'isica Te\'orica y del Cosmos and Centro Andaluz de F\'isica de Part\'iculas Elementales, Facultad de Ciencias, Avda Fuentenueva s/n,
Universidad de Granada, 18071 Granada, Spain}

\emailAdd{\text{jose.beltran@usal.es}}
\emailAdd{\text{dbet@unileon.es}}
\emailAdd{\text{philippe.brax@ipht.fr}}
\emailAdd{\text{bjanssen@ugr.es}}
\emailAdd{\text{pablo.sampedro.fernandez@rai.usc.es}}

\abstract{We study finite size effects on the linear response of spherically symmetric objects in Born-Infeld (BI) electromagnetism and Dirac-Born-Infeld (DBI) scalar field theories. Previous works show that the linear response coefficients for a point-like source vanish for odd multipoles above the dipole, a feature that resembles the vanishing of Love numbers for black holes. This work goes beyond the point-like idealisation and considers a sphere of finite radius. We find that the vanishing of the linear response coefficients ceases as they acquire a correction due to the finite size of the object. This introduces a hierarchy between the even and odd multipoles of the response coefficients determined by the separation of scales between the radius of the sphere and the screening scale of non-linearities. From a phenomenological viewpoint, the hierarchy between the odd and even multipoles would give access to the screening scale and the object's radius by measuring the behaviour of the potentials at infinity.}
\date{\today}

%\keywords{}

\date{\today}
\maketitle
%\newpage
\section{Introduction}

Born-Infeld (BI) electromagnetism \cite{Born:1933pep,Born:1934gh} stands out among the class of non-linear electrodynamics for its exceptional properties. It was originally designed as a completion of Maxwell's electromagnetism so that the electric field of a point-like charged particle remains everywhere regular, including at the position of the particle, thus rendering the energy finite. It was later found that BI-like theories arise naturally in open string theories around D-branes \cite{Fradkin:1985qd,Gibbons:2001gy}. Despite sharing the duality invariance of Maxwell's theory (a property common to a larger class of non-linear electromagnetisms, see e.g. \cite{Gibbons:1995cv,Gaillard:1997rt,Hatsuda:1999ys,Murcia:2025psi}), BI stands out by its causal propagation and the absence of shock waves \cite{Deser:1998wv,Gibbons:2000xe}. It has also been shown that BI can be obtained by imposing certain conditions on the soft limit of scattering amplitudes \cite{Cheung:2018oki}. At a phenomenological level, assuming that the standard photon is described by a BI Lagrangian puts tight constraints on the scale of the non-linearities \cite{Fouche:2016qqj}. Such bounds are less stringent for a dark sector with a dark BI photon. In this context, scenarios of dark matter interacting via a dark BI photon have been explored with interesting effects on structure formation \cite{BeltranJimenez:2020tsl} and even the potential to alleviate the Hubble tension \cite{BeltranJimenez:2020csl,BeltranJimenez:2021imo}.

A closely related theory involving a shift-symmetric scalar field theory is the so-called Dirac-Born-Infeld (DBI) theory. As its BI relative, DBI satisfies some exceptional properties regarding causal propagation, absence of caustics or scattering amplitude soft limits among the general class of scalar field theories \cite{Deser:1998wv,Mukohyama:2016ipl,deRham:2016ged,Cheung:2014dqa}. It can also be obtained from the properties of soft scattering amplitudes \cite{Cheung:2014dqa}. Additionally, DBI can also be singled out by the existence of additional non-linearly realised symmetries \cite{Creminelli:2013xfa,Pajer:2018egx,Grall:2019qof}. Different versions of scalar DBI theories have been used as models of inflation, dark energy, super-fluid dark matter or in the context of classicalisation (see e.g. \cite{Silverstein:2003hf,Martin:2008xw,Copeland:2010jt,Berezhiani:2025maf,Dvali:2010jz} for a non-exhaustive list). 

A common feature of non-linear theories of electromagnetism and shift-symmetric scalar field theories is the existence of a kinetic screening mechanism, also called K-mouflage in the context of scalar fields \cite{Babichev:2009ee,Brax:2012jr}.\footnote{For the specific case of DBI, the screening has been dubbed D-BIonic in \cite{Burrage:2014uwa}.} As a matter of fact, it is this very screening mechanism what lies behind the regularisation of the self-energy of point-like particles. The screening mechanism works due to the non-linearities of the field. Far from the sources (assumed to have compact support), the interactions are sub-dominant so the field takes the usual form of canonical linear theories. As we approach the sources however, the non-linearities become more important until they even dominate the dynamics. The region where this transition happens is called screening surface (or screening radius for a spherically symmetric source).

In previous works \cite{BeltranJimenez:2022hvs,BeltranJimenez:2024zmd}, we have contributed to  revealing  further the exceptional nature of BI and DBI by studying the linear response of point-like objects. For the linear theories (Maxwell electromagnetism and a canonical massless scalar field), a point-like particle does not respond to external stimuli by its own nature. This is essentially due to the lack of self-interactions of the fields. However, for non-linear theories this is no longer the case and a point-like particle can respond to external perturbations. A nice way of understanding it is by considering that the particle together with its screened region behave as a soft object. Upon the action of an external perturbation, the screened volume will be deformed as  the non-linearities are important inside and the external perturbation will interact with the particle's field. This interaction modifies the field generated by the particle thus creating a decaying tail at infinity that characterizes the linear response. The remarkable results found in \cite{BeltranJimenez:2022hvs} and \cite{BeltranJimenez:2024zmd} for BI and DBI respectively is that some multipoles exhibit vanishing linear response, a feature that resembles the vanishing of Love numbers for black holes ~\cite{Binnington:2009bb,LeTiec:2020bos,Charalambous:2021mea}. A similar property has also been observed for analogue black holes \cite{DeLuca:2024nih}.

The present work builds up on \cite{BeltranJimenez:2022hvs,BeltranJimenez:2024zmd} and goes beyond the point-like approximation by considering finite size corrections. In particular, we will study the linear response of a spherical object whose size is smaller than its screening radius. We will see that the vanishing of the linear response coefficients for certain multipoles occurring in the point-like approximation ceases and they acquire corrections due to the finite size of the object. An interesting consequence of such finite size effects is that the separation of scales between the screening radius and the size of the sphere induces a hierarchy between odd and even multipoles for the linear response coefficients. This is phenomenologically appealing as this opens up the possibility of inferring the size of the object and its screening scale from measurements of even and odd multipoles respectively far away from the object itself.

This paper is structured as follows: We begin by briefly reviewing the basics of BI electromagnetism and its close relation to the scalar DBI theory as well as details of their screening mechanism in Sec. \ref{Sec:Thetheories}. We will then solve the static field configuration generated by a sphere of finite size and derive the equations for the static perturbations around it in Sec. \ref{Sec:ScreenedSphere}. After deriving the necessary perturbation equations, we proceed in Sec. \ref{Sec:PolarModes} to solve the perturbation equations for the polar perturbations, which are common to BI and DBI, and to compute the corresponding linear response coefficients. We also review the existence of ladder operators and conserved charges and work out their action on the physical solutions. In Sec. \ref{Sec:AxialModes} we consider the axial perturbations that are genuine of BI but do not exist in DBI. Finally, Sec. \ref{Sec:Discussion} discusses our main results and findings. 

%%%%%%%%%%%%%%%%%%%%%%%%%%%%%%%%%%%%%%%%%%%%%%%%%%
\section{The theories}
\label{Sec:Thetheories}
%%%%%%%%%%%%%%%%%%%%%%%%%%%%%%%%%%%%%%%%%%%%%%%%%%

In this Section we will review the basics of BI electromagnetism and its DBI scalar counterpart. This will not only serve to set-up the relevant equations for the core of the paper, but it will also permit us to establish the close relation between both theories.

%%%%%%%%%%%%%%%%%%%%%%%%%%%%%%%%%%%%%%%%%%%%%%%%%%
\subsection{Born-Infeld electromagnetism}
\label{Sec:BI}
%%%%%%%%%%%%%%%%%%%%%%%%%%%%%%%%%%%%%%%%%%%%%%%%%%

The Lagrangian that describes BI electromagnetism is given by:
\be
\Lag_{\rm BI}=\Lambda^4\left(1-\sqrt{1-\frac{2Y}{\Lambda^4}-\frac{Z^2}{\Lambda^8}}\right)\,,
\label{eq:BIlagrangian}
\ee
where $\Lambda$ is the scale at which non-linearities become important and $Y=-\frac{1}{4}F_{\mu \nu}F^{\mu \nu}$ and $Z=-\frac{1}{4}F_{\mu \nu}\tilde{F}^{\mu \nu}$ are the two independent Lorentz invariants in four dimensions constructed out of the field strength $F_{\mu\nu}=\partial_\mu A_\nu-\partial_\nu A_\mu$ of the vector potential $A_\mu$ and its dual $\tilde{F}^{\mu\nu}=\frac12\epsilon^{\mu\nu\alpha\beta}F_{\alpha\beta}$. In terms of the electric and magnetic components of the field strength, these two Lorentz invariants can be expressed as $Y=\frac{1}{2}(\vec{E}^2-\vec{B}^2)$ and $Z=\vec{E}\cdot \vec{B}$. In this work we are interested in static configurations so we will neglect any time-dependence. In that case, the field equations reduce to
\be
\nabla\cdot\vec{D}=\rho,\quad\nabla\times\vec{H}=0,
\ee
where $\rho$ is a static source and we have defined
\bea
\vec{D}=\frac{\partial \Lag_{\text{BI}}}{\partial\vec{E}}=\partial_Y\Lag_{\text{BI}}\vec{E}+\partial_Z\Lag_{\text{BI}}\vec{B}\,,\label{eq:defD}\\
\vec{H}=-\frac{\partial \Lag_{\text{BI}}}{\partial\vec{B}}=\partial_Y \Lag_{\text{BI}}\vec{B}-\partial_Z\Lag_{\text{BI}}\vec{E}\,.
\label{eq:defH}
\eea
Besides these equations, we also have the Bianchi identities $\nabla_\mu\tilde{F}^{\mu\nu}=0$ which, for our static configuration, read
\be
\nabla\times\vec{E}=0,\quad\nabla\cdot \vec{B}=0.
\ee
As usual, the first Bianchi identity for static configurations tells us that the electric field can be written as $\vec{E}=-\nabla\phi_e$, with $\phi_e$ the electric potential. This is an off-shell identity. Furthermore, the dynamical equation for static configurations $\nabla\times\vec{H}=0$ additionally imposes that $\vec{H}$ also derives from a (pseudo-)scalar potential $\vec{H}=\nabla\psi$, although this time it is an on-shell relation. Thus, static configurations can be fully characterised in terms of two scalar potentials, namely, $\phi_e$ and $\psi$, that we will refer to as polar and axial sectors respectively. This is all we will need for our purposes in this work, so we will not provide more details about the properties and remarkable features of BI electromagnetism here. Instead, let us see how BI is related to its scalar relative.

\subsection{DBI}
The Lagrangian for the (shift-symmetric) scalar DBI can be written as
\be
\Lag_{\text{DBI}}=\Lambda^4\left(\sqrt{1+\frac{2X}{\Lambda^4}}-1\right),
\label{eq:LDBI}
\ee
with $X=-\frac12(\partial\phi)^2$ and $\Lambda$ some energy scale.\footnote{This scale is not related to the scale introduced for BI, but we will use the same symbol to avoid unnecessary redundant notations. This will also facilitate a unified treatment.} The sign of $\Lambda^4$ has been fixed so that the theory exhibits a screening mechanism \cite{Burrage:2014uwa}. The opposite sign is the one relevant for cosmological applications where the scalar field takes a time-like profile $\langle\phi\rangle=\phi(t)$, but we are not interested in those in this work. Instead, we are interested in static configurations so our background takes the form $\langle\phi\rangle=\phi(\vec{x})$ and we have $X=-\frac12\vert\nabla\phi\vert^2$. The DBI Lagrangian for these static configurations reduces to
\be
\Lag^{\text{static}}_{\text{DBI}}=\Lambda^4\left(\sqrt{1-\frac{\vert\nabla\phi\vert^2}{\Lambda^4}}-1\right).
\ee
We can now compare this expression to the BI Lagrangian for a static and purely electric configuration
\be
\Lag^{\text{static}}_{\text{BI}}(\vec{B}=0)=\Lambda^4\left(1-\sqrt{1-\frac{\vert\nabla\phi_e\vert^2}{\Lambda^4}}\right),
\ee
so we see that, up to an irrelevant constant and a global sign, both Lagrangians coincide. As a matter of fact, the equation for the DBI field in a static configuration is given by
\be
\nabla\cdot\left(\Lag_{\text{DBI}}'\nabla\phi\right)=\rho,
\ee
which exactly coincides with that of BI with $\vec{B}=0$ and upon identifying $\vec{D}\to\Lag_{\text{DBI}}'\nabla\phi$. It is important to notice that this equivalence holds at full non-linear order as long as we maintain staticity and a trivial magnetic sector.\footnote{It may be worth noticing that having a purely electric configuration, i.e., vanishing $\vec{B}=0$ is a consistent truncation in BI and, as a matter of fact, in parity invariant non-linear electrodynamics in general.} Thus, from now on we will treat both DBI and the electric sector of BI together. In particular, we will write $\vec{E}$ to refer to either $\nabla\phi$ or $\nabla\phi_e$ indistinctively (unless otherwise stated) and we will call it {\it electric field} for both cases, although for DBI it does not really represent an electric field. However, in both cases $\vec{E}$ will describe the mediated force. We will also use the notation $Y=\vec{E}^2/2$ and will describe the Lagrangians of BI and DBI by means of the common function
\be
\mK(Y)=\sqrt{1-\frac{2Y}{\Lambda^4}}=\sqrt{1-\frac{\vec{E}^2}{\Lambda^4}}.
\ee
The interesting common feature of both theories is the presence of a screening mechanism that will play an important role in our study. Thus, before proceeding to the main goal of this paper,  it will be convenient to review briefly the basics of the screening mechanism present in BI and DBI for static configurations.

\subsection{Screening}
Let us consider a source with compact support. For this source distribution, the field is expected to decay asymptotically and, in particular, we expect to have $\vec{E}^2\ll \Lambda^4$. In this regime, the dynamics are governed by the linear term in an expansion of the Lagrangian, and the monopolar contribution will dominate (an approximate SO(3) symmetry arises) so the field will decay as $\phi\sim 1/r$.  The scalar field profile then grows as we approach the source and, at some point, we will reach a regime where $\vec{E}^2\simeq\Lambda^4$. When this happens, the non-linearities will not only become relevant, but they will in turn govern the behaviour of the field. This region is what is called the screened region \cite{Babichev:2009ee,Brax:2012jr,Brax:2014gra}. If the monopole is still the dominant contribution, the region will be a sphere to a good approximation and the radius of such a sphere is called the screening radius $\rs$. If the multipolar expansion breaks down before entering the strong field regime with $\vec{E}^2\simeq \Lambda^4$, so that higher order multipoles are no longer negligible, the screened region will have a more complicated geometry. An important feature of BI and DBI is that the square root structure of the Lagrangian bounds the electric field to be $\vec{E}^2\lesssim\Lambda^4$. In fact, the static solution saturates to $\vert\vec{E}\vert\simeq\Lambda^2$ throughout the entire screened region. 

The BI and DBI theories are clearly non-renormalizable theories so it may be worrisome that the screening relies on having $E\simeq\Lambda^2$ that could cause higher order operators to become important. However, both BI and DBI (can) make sense as EFTs within the screening region because the breakdown of the perturbative expansion occurs when $\partial E\sim \Lambda E$ \cite{deRham:2014wfa,Brax:2016jjt}. This permits to have a classical regime with $E\sim\Lambda^2$ and quantum corrections under control as long as $\partial \ln E\ll\Lambda$. Having this issue under control however does not guarantee the absence of problematic features. For a point like particle, the perturbations around the screened solution have some worrisome properties. For the BI theory, the perturbations propagate subliminally inside the screened region, which is a pleasant property that in turn reflects the compliance of BI with positivity bounds \cite{Adams:2006sv}. However, the propagation speed inside the screened region scales as $r^4$, which means that it becomes arbitrarily small as we approach the origin. This may cause the appearance of a strong coupling problem because, for instance, spatial gradients of the perturbations are arbitrarily easy to excite sufficiently close to the origin. For the opposite sign DBI we are considering, the propagation speed becomes superluminal, which conflicts with standard positivity constraints \cite{Adams:2006sv}, and becomes arbitrarily large as we approach the origin. As a matter of fact, this dual behaviour occurs because the propagation speeds in BI and DBI are one the inverse of the other $c_{\text{DBI}}^2=1/c_{\text{BI}}^2$ and this can be understood because $E$ represents kinetic energy in BI but potential energy (associated to spatial gradients) in DBI. These potential problems are exacerbated in the case of a point-like source where the screened region extends to the whole region surrounding the particle so the pathologies can become arbitrarily severe. This means that the regime of validity of the solutions might break down before than expected. Considering a finite size object instead of a point-like particle eases some of these potential problems and this is what we will consider in the subsequent sections.

\section{Screened sphere}
\label{Sec:ScreenedSphere}

In this Section we will analyse the field profile generated by a sphere of radius $r_0$ with a constant density $\rho_0$. We will first compute the background configuration and later we will obtain the equations for static perturbations around such a background. 

\subsection{Background profile}

\begin{figure*}[hb]
\centering
\includegraphics[width=0.75\linewidth]{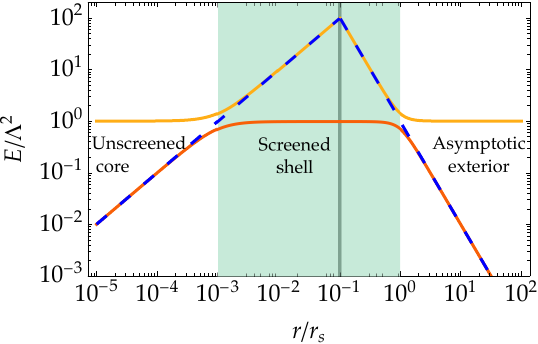}
\caption{Electric field profile for a sphere in BI and DBI (solid-orange). The case of standard linear theories, i.e., Maxwell electromagnetism and a canonical massless scalar field, is also shown in dashed line for comparison. We have indicated the radius of the sphere (that we choose as $r_0=10^{-1}\rs$) with a vertical line, and the screened shell is shown in light green shading. Far from the sphere we observe the standard $1/r^2$ decay. Inside the screened shell, the field saturates to $\Lambda^2$ and is suppressed with respect to the linear theories. In this screened shell, the derivative of the non-linear function $\mK_Y$ (yellow line) coincides with the field of the linear theory, as expected because $1/\mK_Y$ is precisely the screening factor and the background field is constant in that region. Finally, the screening ceases in the inner unscreened core and we recover again the behaviour of the linear theories.}
\label{fig:Eback}
\end{figure*}

The background electric field for a static and spherically symmetric source is governed by the equation
\be
\nabla\cdot\vec{D}=\rho(r),
\ee
which can be straightforwardly solved for our spherically symmetric object of radius $r_0$ as
\be
\vec{D}=\frac{Q(r)}{4\pi r^2}\hat{r},
\label{eq:Dconfig}
\ee
where $\hat{r}$ is a position unit vector and 
$Q(r)\equiv4\pi\int\dd r \,r^2\rho(r)$ is the charged enclosed by a sphere of radius $r$. Outside the sphere, the charge is constant and coincides with the total charge of the object $Q_0\equiv\tfrac{4\pi}{3}\rho_0 r_0^3$. Inside the sphere we have instead the charge profile $Q(r)=Q_0(r/r_0)^3$. The constitutive relation between $\vec{D}$ and $\vec{E}$ for BI and DBI is given by
\be
\vec{D}=\frac{\vec{E}}{1-\vec{E}^2/\Lambda^4},
\ee
that can be inverted to obtain
\be
\vec{E}=\frac{\vec{D}}{\sqrt{1+\vec{D}^2/\Lambda^4}}.
\ee
Using the solution for $\vec{D}$ given in \eqref{eq:Dconfig}, we can finally obtain the electric field profile that can be expressed as
\be
\vec{E}=\frac{1}{\sqrt{1+\left(\tfrac{Q(r)}{4\pi\Lambda^2 r^2}\right)^2}}\,\frac{Q(r)}{4\pi r^2}\,\hat{r}.
\ee
The profile of this electric field is shown in Fig. \ref{fig:Eback}, where we can distinguish the following three regimes:
\begin{itemize}

\item {\bf{Asymptotic (unscreened) exterior}} $r\gg\sqrt{\tfrac{Q_0}{4\pi}}\Lambda^{-1}\equiv r_s$. In this region, the non-linearities are negligible and we obtain the standard profile for a charge $Q_0$. The screening radius $\rs$ determines when the non-linearities become important. In this work we will assume that such a radius is well-outside the sphere.
    
\item \bf{Screened shell} $r_0^3/\rs^2\ll r\ll\rs$. In this region the non-linearities become important and the field saturates to $E\simeq\Lambda^2$.

\item \bf{Unscreened core} $r\ll r_0^3/\rs^2$. This region appears because, as we enter the sphere, the total charge enclosed by spheres of a given radius is reduced by a factor $r^3/r_0^3$ and this makes the non-linearities become less important as we go deeper in the sphere. When the total enclosed charge is such that $\tfrac{Q(r)}{4\pi \Lambda^2 r^2}\simeq 1$, the non-linearities will no longer be relevant and the screening will cease working. This occurs at a radius $r_0^3/\rs^2$ that we will call inner screening radius and characterises the size of the inner unscreened core. This region does not exist for the point-like particle for which $r_0\to0$.
\end{itemize}

The described field configuration for a sphere in BI electromagnetism does not seem to have been analysed in the literature, probably because the main interest is in studying the regularisation of point-like sources. A similar field configuration has been noticed for polynomial shift-symmetric scalar field theories, $P(X)$, in e.g. \cite{terHaar:2020xxb,Lara:2022gof} (see also \cite{rodriguez2024astrophysical} for the explicit DBI case). Interestingly, the importance of requiring regularity at the centre of the object was emphasized in \cite{terHaar:2020xxb}, a condition that is automatically satisfied for BI-like theories. Furthermore, another advantage of BI-like theories is the existence of a single branch of solutions, while for e.g. polynomial theories, there are several branches and identifying the physical one connecting a regular solution in the interior with a screened solution is not always a trivial task. Since our focus is on BI-like theories, we will not need to deal with these issues, but it would be interesting to explore further these advantageous features of BI-like theories.

Having properly characterised the background field profile, we are ready to obtain the perturbation equations that will allow us to compute the linear response coefficients for the sphere.

\subsection{Static perturbations}
The detailed derivation of the equations for static perturbations in BI and DBI can be found in e.g.~\cite{BeltranJimenez:2022hvs,BeltranJimenez:2024zmd}. Here we will not repeat the full derivation, but we will simply indicate the main steps and quote the relevant perturbation equations. For the polar perturbations in BI and DBI perturbations (which are governed by the same equations), the starting point is
\be
\nabla\cdot\delta \vec{D}=0,
\label{eq:eqdeltaD}
\ee
where 
\be
\delta\vec{D}=\mK_Y\delta\vec{E}+\mK_{YY}(\vec{E}\cdot\delta\vec{E})\vec{E},
\ee
with $\mK_Y=\partial_Y\mK$ and $\mK_{YY}=\partial_{YY}\mK$. Taking into account that the background field is radial and introducing the convenient variable
\be
\Phi\equiv \frac{r^2\mK_Y}{\cBI^2}\partial_r\phi,
\ee
with\footnote{The parameter $\cBI^2$ describes the propagation speed of perturbations for BI electromagnetism. One of the remarkable properties of BI is that this propagation speed is the same for both polarisations, i.e., there is no birefringence and the propagation admits a geometrical description, which turn out to describe a wormhole geometry \cite{Jimenez:2024npb}. For DBI, the propagation speed of perturbations is $c^2_{\text{DBI}}(r)=1/\cBI^2(r)$. Thus, while the BI perturbations are subluminal for screened backgrounds, the DBI exhibits superluminalities.}
\be
\cBI^2=1+\frac{2Y\mK_{YY}}{\mK_{Y}}
\ee
we can write \eqref{eq:eqdeltaD} as
\be
\partial_r\Phi+\mK_Y\nabla^2_\Omega\phi=0.
\ee
We can take a radial derivative of this equation and use the definition of $\Phi$ to obtain the equation
\be
\Phi''-\partial_r\ln\mK_Y\Phi'+\frac{\mK_Y}{r^2\left(\mK_Y+2Y\mK_{YY}\right)}\nabla^2_\Omega\phi=0.
\ee
If we now decompose into spherical harmonics as $\Phi=\sum_{\ell,m}\sqrt{\mK_Y}\Phi_\ell(r) Y_{\ell,m}$, the resulting equation can be written as
\be
\Phi_\ell''-m_P^2\Phi_\ell=0,
\ee
with 
\be
m_P^2=\frac{\cBI^2\ell(\ell+1)}{r^2}+\frac14(\partial_r\ln\mK_Y)^2-\frac12\partial^2_r\ln\mK_Y\,.
\label{eq:polarmass}
\ee

The axial perturbations for BI (which do not exist for DBI) can be obtained following a similar procedure (we refer to \cite{BeltranJimenez:2022hvs} for the details). As we have explained above, for static configurations we have the on-shell relation $\vec{H}=\nabla\psi$ with $\psi$ the axial scalar potential. Using the constitutive relation \eqref{eq:defH} to express $\delta\vec{B}$ in terms of $\psi$ and the Bianchi identity $\nabla\cdot\delta\vec{B}=0$, we can obtain the equation
\be
\partial_r\Psi+\frac{1}{\mK_Y}\nabla_\Omega^2\psi=0\,,
\ee
where we have defined
\be
\Psi\equiv\frac{r^2}{\mK_Y\cBI^2}\partial_r\psi.
\ee
We can see that this equation is the same as for the polar sector upon the substitution $\mK_Y\to1/\mK_Y$, so we only need to make that replacement in the equation of the polar modes to obtain the axial perturbation equations. More explicitly, we will have
\be
\Psi_\ell''-m_A^2\Psi_\ell=0,
\ee
with 
\be
m_A^2=\frac{\cBI^2\ell(\ell+1)}{r^2}+\frac14(\partial_r\ln\mK_Y)^2+\frac12\partial^2_r\ln\mK_Y\,,
\label{eq:polarmass}
\ee
where we have used the multipolar decomposition $\Psi=\sum_{\ell,m}\frac{1}{\sqrt{\mK_Y}}\Psi_\ell(r) Y_{\ell,m}$.
These perturbation equations are linked to supersymmetric quantum mechanics (see \cite{BeltranJimenez:2022hvs} for details). 

Equipped with the perturbation equations, we can proceed to the main focus of this work that consists in computing the linear response of the sphere and, for that, we need to solve the obtained perturbation equations with appropriate boundary conditions.

\section{Polar modes}
\label{Sec:PolarModes}

In this Section we will compute the linear response coefficients of the sphere under some external perturbation. We will proceed by solving the equations for the perturbations both inside and outside the sphere and then matching both solutions at the surface of the sphere with the appropriate matching conditions that we will also obtain. In particular, we will see that we need to impose a discontinuity in the first derivatives. As a notational comment, we will use the screening scale $\rs$ to normalise radial variables and masses. In particular, we will use the radial variable $x\equiv r/\rs$ from now on so the radius of the sphere will be $x_0=r_0/\rs$. We will also use the convenient parameter $\alpha\equiv x_0^{-3}$.

\subsection{Exterior}

The effective mass for the polar perturbations outside the sphere reads:
\be
\rs^2m_{\text{P,out}}^2=\left[\ell(\ell+1)-\frac{5}{1+x^4}\right] \frac{x^2}{1+x^4},
\ee
which is of course the same as that of a point-like particle. We can then follow \cite{BeltranJimenez:2022hvs} and introduce the variable $z=-x^4$ together with the field redefinition $\tilde{\Phi}_\ell=(1+x^4)^{1/4}\Phi_\ell$ to write the equation as the following hypergeometric equation:
\be
z(1-z)\frac{\dd^2\tilde{\Phi}^{\text{out}}_\ell}{\dd z^2}+\frac{3-z}{4}\frac{\dd\tilde{\Phi}^{\text{out}}_\ell}{\dd z}+\frac{\ell(\ell+1)-2}{16}\tilde\Phi^{\text{out}}_\ell=0,
\label{eq:polarhypereq}
\ee
that can be solved in terms of hypergeometric functions so that the solution in the original variables is
\be
\Phi^{\text{out}}_\ell=\frac{1}{(1+x^4)^{1/4}}\bigg[A^{\text{out}}_{\ell}  \;_2F_1\left(-\tfrac{\ell+2}{4},\tfrac{\ell-1}{4},\tfrac{3}{4}, -x^4\right)+B^{\text{out}}_{\ell}\;x  \;_2F_1\left(-\tfrac{\ell+1}{4},\tfrac{\ell}{4},\tfrac{5}{4},-x^4\right)\bigg],
\label{eq:gensolpolout}
\ee
with $A^{\text{out}}_\ell$ and $B^{\text{out}}_\ell$ integration constants to be fixed by boundary conditions. One boundary condition will be set by matching the asymptotic growing tail with the amplitude of the external perturbation. The other boundary condition will be fixed by matching to the solution inside the sphere. The solutions inside the sphere will be required to be regular at the origin and, as we will show in the following, this uniquely selects the second boundary condition. Then, let us turn to the solutions inside the sphere.

\subsection{Interior}

The effective mass inside the sphere is given by
\be
\rs^2 m_{\text{P,in}}^2=\frac{4\ell(\ell+1)+2\big[2\ell(\ell+1)-1\big]\alpha^2 x^2+3\alpha^4 x^4}{4x^2(1+\alpha x^2)^2},
\label{eq:hyperpolin}
\ee
where we recall that $\alpha=x_0 ^{-3}$. For $x\ll1$, this effective mass reduces to $m_{\text{P,in}}^2\simeq\tfrac{\ell(\ell+1)}{x^2}$ which is the Maxwellian behaviour. This reflects the fact that the screening mechanism ceases working sufficiently close to the origin. We can introduce the variable $z\equiv-\alpha^2 x^2$ and the field redefinition $\tilde{\Phi}_\ell\equiv x^\ell (1+\alpha^2x^2)^{1/4}\Phi_\ell$ to obtain the equation
\be
z(1-z)\frac{\dd^2\tilde{\Phi}^{\text{in}}_\ell}{\dd z^2}+\frac{1+2\ell(z-1)}{2}\frac{\dd\tilde{\Phi}^{\text{in}}_\ell}{\dd z}-\frac{\ell(\ell+2)}{4}\tilde\Phi^{\text{in}}_\ell=0,
\label{eq:hyperpolarin}
\ee
which is again a hypergeometric equation. Going back to our original variable, the solutions can be written as
\begin{align}
\Phi^{\text{in}}_\ell=&\frac{\sqrt{\alpha}}{\left(1+\alpha^2x^2\right)^{1/4}} \Big[A^{\text{in}}_\ell x^{-\ell}\ _2F_1\left(-\tfrac{2+\ell}{2},-\tfrac{\ell}{2},\tfrac{1-2\ell}{2},-\alpha^2x^2\right)\nonumber\\&
+B^{\text{in}}_\ell\alpha ^{2\ell+1} x^{\ell+1} \ _2F_1\left(\tfrac{\ell-1}{2},\tfrac{\ell+1}{2},\tfrac{3+2\ell}{2},-\alpha^2x^2\right)  \Big].
\end{align}
Now that we have the solution inside the sphere, we can proceed to impose the second boundary condition that will be fixed by requiring regularity of the perturbation at the origin. Let us notice at this point that this is a valid boundary condition with no subtleties because the source is perfectly smooth there. This contrasts with the case of a point-like object for which imposing regularity at the origin might not be justified in general precisely because of the singular nature of the source. This happens for the standard Maxwell theory and a canonical massless scalar field for instance. BI and DBI are remarkable in this respect because the background field remains regular even at the position of the particle so those theories naturally permit to require regularity of the perturbations at the origin and this property distinguishes them from other non-linear electromagnetisms and shift-symmetric K-essence theories. Returning to our finite sphere, we have the following asymptotic expansion of the solution near the origin:
\be
\Phi_\ell^{\text{in}}(x\to0)\simeq \sqrt{\alpha}A^{\text{in}}_\ell x^{-\ell}-\alpha^{2\ell+\frac{3}{2}}B^{\text{in}}_\ell x^{\ell+1}\,,
\ee
so we need to impose $A^{\text{in}}_\ell=0$ for $\ell>0$. For $\ell=0$, the effect of the perturbation can be simply absorbed  into the background by a renormalisation of the charge. Thus, the physical solution inside the sphere is
\begin{align}
\Phi^{\text{in}}_\ell=&B^{\text{in}}_\ell\alpha ^{2\ell+\frac{3}{2}}\frac{ x^{\ell+1}}{\left(1+\alpha^2x^2\right)^{1/4}}  \ _2F_1\left(\tfrac{\ell-1}{2},\tfrac{\ell+1}{2},\tfrac{3+2\ell}{2};-\alpha^2x^2\right).
\label{eq:phiingreg}
\end{align}
Now that we have the solutions inside and outside the sphere, we can proceed to computing the linear response coefficients.

\subsection{Polar linear response}

\begin{figure}[ht]
\includegraphics[width=0.49\linewidth]{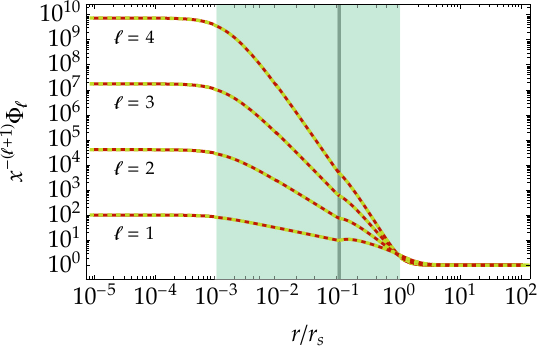}
\includegraphics[width=0.49\linewidth]{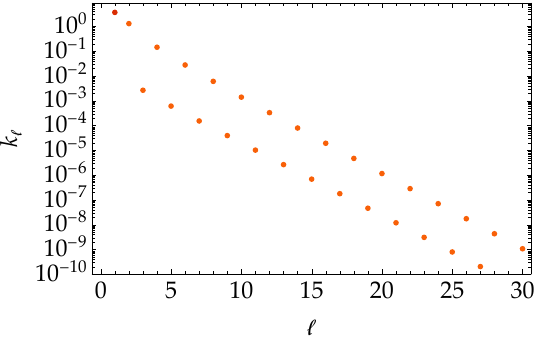}
\caption{
In the left panel we show the solutions for a few multipoles normalised to the amplitude of the external perturbation for a sphere of radius $x_0=10^{-1}$. We show the exact numerical solution (solid) together with the analytical solution (dashed) obtained by matching at the surface of the sphere taking into account the discontinuity in the first derivative. We can clearly observe how the perturbation grows as it transits through the screened shell (shaded region) and it has the usual behaviour $\sim x^{\ell+1}$ in the asymptotic region and the unscreened core. In the right panel we plot the linear response coefficients where we can see the hierarchy between odd and even multipoles explained in the main text.
}
\label{Fig:polar}
\end{figure}

In the asymptotic region $x\to\infty$, the solution is expected to have, generically, a growing mode describing the external perturbation and a decaying tail that accounts for the response of the object. If we expand the outer solution in this asymptotic region we find
\begin{align}
\Phi_\ell^{\text{out}}(x)\simeq & \,\Gamma\left(-\tfrac{2\ell+1}{4}\right)\left[ \frac{\Gamma(\tfrac{3}{4})}{\Gamma(-\tfrac{\ell+2}{4})\Gamma(\tfrac{4-\ell}{4})}A^{\text{out}}_\ell+\frac{\Gamma(\tfrac{5}{4})}{\Gamma(-\tfrac{\ell+1}{4})\Gamma(\tfrac{5-\ell}{4})}B^{\text{out}}_\ell\right]x^{-\ell}\\ &+\Gamma(\tfrac{2\ell+1}{4})\left[\frac{\Gamma(\tfrac{3}{4})}{\Gamma(\tfrac{\ell-1}{4})\Gamma(\tfrac{5+\ell}{4})}A^{\text{out}}_\ell+\frac{\Gamma(\tfrac{5}{4})}{\Gamma(\tfrac{\ell}{4})\Gamma(\tfrac{6+\ell}{4})}B^{\text{out}}_\ell\right]x^{\ell+1}\,,
\end{align}
that confirms the presence of a growing and a decaying tail. The linear response is then defined as the coefficient of the decaying tail over the coefficient of the growing tail:\footnote{In this work we will work with the dimensionless linear response coefficients $k_\ell$ unlike in \cite{BeltranJimenez:2022hvs,BeltranJimenez:2024zmd} where dimensionfull coefficients, denoted as $\alpha_\ell$ (not to be confused with $\alpha\equiv x_0^{-3}$ here), are employed instead. The difference simply amounts to restoring the appropriate units using the screening radius $\alpha_\ell=k_\ell\rs^{2\ell+1}$.\label{FN:klalphal}}
\be
k_\ell\equiv\frac{\Gamma\left(-\tfrac{2\ell+1}{4}\right)\left[ \frac{\Gamma\left(\tfrac{3}{4}\right)}{\Gamma\left(-\tfrac{\ell+2}{4}\right)\Gamma\left(\tfrac{4-\ell}{4}\right)}A^{\text{out}}_\ell+\frac{\Gamma\left(\tfrac{5}{4}\right)}{\Gamma\left(-\tfrac{\ell+1}{4}\right)\Gamma\left(\tfrac{5-\ell}{4}\right)}B^{\text{out}}_\ell\right]}{\Gamma\left(\tfrac{2\ell+1}{4}\right)\left[\frac{\Gamma\left(\tfrac{3}{4}\right)}{\Gamma(\tfrac{\ell-1}{4})\Gamma\left(\tfrac{5+\ell}{4}\right)}A^{\text{out}}_\ell+\frac{\Gamma\left(\tfrac{5}{4}\right)}{\Gamma\left(\tfrac{\ell}{4}\right)\Gamma\left(\tfrac{6+\ell}{4}\right)}B^{\text{out}}_\ell\right]}\,.
\ee
In order to compute this quantity, we need to know the values of the integration constants $A^{\text{out}}_\ell$ and $B^{\text{out}}_\ell$, which are determined by boundary conditions. The condition of having a regular solution at the center $A^{\text{in}}_\ell=0$ can be translated to the outer solution by matching the interior and exterior solutions at the surface of the object $x=x_0$. In order to match properly both solutions we first note that the screening factor $\mK_Y$ presents a discontinuity in its first derivative at the surface of the sphere. Since the effective mass depends on the radial derivatives of $\log\mK_Y$ up to second order, there will be  a Dirac delta function localized at the surface of the sphere and this will generate a discontinuity in the first derivative  of $\Phi$ at $x=x_0$.\footnote{This discontinuity occurs for a hard sphere with a sharp edge. If we instead consider a sphere described by a density profile $\rho(r)$ that smoothly connects the interior where $\rho_{\text{int}}=\rho_0$ with the exterior where $\rho_{\text{ext}}\ll\rho_0$, the first derivative of $\phi$ will be continuous everywhere. The jump will effectively capture the behaviour of $\Phi$ within the transition shell separating the interior and the exterior of the sphere.} To compute this discontinuity, we use that we can write $\mK_Y$ inside the screened region as follows
\be
\frac{\mK_Y(x)}{\mK_Y(x_0)}=\begin{cases}
x/x_0 &  x < x_0, \\[0.75em]
(x/x_0)^{-2} &  x > x_0.
\end{cases}
\ee
From this expression we can then compute its first radial derivative to obtain the following relation:
\be
(\log\mK_Y)'=\frac{1}{x}\Big[1-3\Theta(x-x_0)\Big],
\ee
with $\Theta(x)$ the Heaviside step function. By taking another derivative we then obtain
\be
(\log\mK_Y)''=-\frac{1}{x}(\log\mK_Y)'-\frac{3}{x}\delta(x-x_0).
\ee
With this result we can compute the discontinuity in the first derivative $(\Delta\Phi')_{x_0}$ in the usual way by integrating the equation in an infinitesimal interval around the surface of the sphere
\be
(\Delta\Phi_\ell')_{x_0}=\lim_{\epsilon\to0}\int_{x_0-\epsilon}^{x_0+\epsilon}\mP^2(x)\Phi_\ell(x)\dd x=\frac{3}{2x_0}\Phi_\ell(x_0),
\ee
where we have used the expression for the effective mass \eqref{eq:polarmass} and we have taken into account that the propagation speed is continuous. Since the solution is continuous, we can use either the interior or the exterior solution to evaluate $\Phi_\ell(x_0)$ in the right hand side of the above expression. Thus, our matching conditions will be
\be
\Phi_\ell^{\text{in}}(x_0)=\Phi_\ell^{\text{out}}(x_0),\qquad \left[\frac{\dd\Phi_\ell^{\text{out}}}{\dd x}-\frac{\dd\Phi_\ell^{\text{in}}}{\dd x}\right]_{x_0}=\frac{3}{2x_0}\Phi_\ell(x_0)\,.
\ee
We have explicitly checked that these matching conditions correctly reproduce the numerical solutions (see Fig. \ref{Fig:polar}). The obtained matching conditions allow us to express all integration constants in terms of one of them. This remaining constant is then fixed by the amplitude of the external perturbation, but this is irrelevant for our purposes because it factors out from the exterior solution and the linear response coefficient $k_\ell$ is oblivious to it, as it should. The resulting expression is very contrived and not particularly illuminating so we will refrain ourselves from giving it here and directly show their numerical evaluation in Fig. \ref{Fig:polar} for a specific case. A more illuminating approach is to perform an expansion for a small radius of the spherical object $x_0\ll1$, which is the interesting case for us because we assume the sphere is well within its screening radius. Otherwise, the strong field regime would never be reached. In this limit we can write the linear response coefficients as
\be
k_\ell \simeq k^{(0)}_\ell+k^{(1)}_\ell x_0,
\ee
with
\bea
k^{(0)}_\ell&=&\frac{\Gamma\left(-\frac{2\ell+1}{4}\right)\Gamma\left(\frac{\ell}{4}\right)\Gamma\left(\frac{\ell+6}{4}\right)}{\Gamma\left(-\frac{\ell+1}{4}\right)\Gamma\left(\frac{5-\ell}{4}\right)\Gamma\left(\frac{2\ell+1}{4}\right)},\\
k^{(1)}_\ell&=&\frac{2^{-4\ell+3/2}(\ell+2)\Gamma\left(-\tfrac{1}{4}\right)\Gamma\left(\tfrac{3-2\ell}{4}\right)\Gamma\left(\tfrac{\ell+4}{2}\right)\Gamma\left(2\ell\right)}{3\ell\,\Gamma\left(\tfrac{5}{4}\right)\Gamma\left(\tfrac{\ell+1}{2}\right)\Gamma\left(\tfrac{2\ell+3}{2}\right)\Gamma\left(\tfrac{2\ell+1}{4}\right)}\sin\left[\frac{(2\ell+1)\pi}{4}\right]\,.
\eea
The leading order contribution $k^{(0)}_\ell$ pleasantly recovers the result for the point like-particle \cite{BeltranJimenez:2022hvs,BeltranJimenez:2024zmd}. An important property of this contribution is that it vanishes for odd multipoles above the dipole, i.e. $k^{(0)}_{2n+1}=0$ for $n=1,2,3,.\cdots$ and a detailed account of this property can be found in \cite{BeltranJimenez:2022hvs,BeltranJimenez:2024zmd}. Our result here shows that the linear response coefficients for those multipoles acquire a correction due to the finite size of the sphere since $k^{(1)}_\ell$ does not vanish for any multipole. Thus, the vanishing of the linear response is tightly tied to the point-like nature of the object. It is worth noticing again that the point-like particle does have special properties within BI and DBI theories because the background profile remains finite and this makes it natural to impose regular boundary conditions even for point-like particles that have a singular density distribution. The finite size sphere does not encounter this problem because the density distribution is perfectly regular everywhere. However, it was not clear a priori that going to the point-like limit from the finite sphere would lead to sensible results given the generally pathological nature of this limit. The fact that BI and DBI comply to provide sensible results originate from their mentioned remarkable features.

From an observational viewpoint, it is advantageous that the odd and even multipoles have different leading orders. Thus, by measuring the linear response of odd multipoles we can directly measure the screening radius scales (see footnote \ref{FN:klalphal}), which is the leading order, while the even multipoles will provide a direct measurement of the size of the object. The hierarchy between the two multipole series can be clearly seen in Fig. \ref{Fig:polar}. This feature roots in the fact that the point-like particle has vanishing linear response for odd multipoles above the dipole.

\subsection{An inner linear response}

So far we have focused on the standard linear response coefficients as measured at infinity and this is related to the deformation of the outer screening surface. In our screened sphere, however, we can also ask the question what is the deformation of the inner screening sphere and, hence, define another set of linear response coefficients as measured in the asymptotically inner region well-inside the unscreened core where the linear theory is also recovered.\footnote{From the perspective of the dielectric medium analogue, the question relates to how the electric field is modified in the centre of a spherical shell with some inhomogeneous electric permittivity.} This question may be pertinent for instance in the scenarios constructed in \cite{BeltranJimenez:2020tsl} where dark matter is charged under a dark BI electromagnetic interaction. In this situation, the inner linear response would be related to how objects (e.g. galaxies/dark matter haloes) well inside a dark matter cluster would react to perturbations outside the cluster. Another situation where this may be relevant is for astrophysical objects with a screening mechanism, where the interior core would also react to external perturbations for screened objects. Although these considerations serve as motivation for studying the inner linear response, we do not aim at providing a full characterisation for said scenarios, but we simply want to point out its potential relevance and show how the field is modified in the unscreened core with respect to its behaviour for the linear theories (Maxwell and the massless canonical scalar).

The effective mass for the linear theory is given by $m_\text{P}^2=\tfrac{\ell(\ell+1)}{r^2}$ and the general solution is
\be
\Phi_\ell^{\text{lin}}=A^{\text{lin}}_\ell x^{-\ell}+B^{\text{lin}}_\ell x^{\ell+1},
\ee
that is valid everywhere due to the linearity of the equations. For $\ell\neq0$, regularity at the origin requires $A^{\text{lin}}_\ell=0$. For $\ell=0$, we obtain a constant mode that can be absorbed into the background charge, so we will only consider $\ell>0$. In the non-linear (D)BI theory, the perturbations will have the same general solution as the linear theory inside the unscreened core and the decaying tail will also be removed by regularity at the center. However, the non-linearities induce a correction to the mode $x^{\ell+1}$ generated by the interactions of the background field with the external perturbation as it goes through the screened shell. This deformation is what defines the inner linear response of the object to the external stimuli. The solution inside the sphere has already been obtained in \eqref{eq:phiingreg} and we can expand it near the origin (well inside the screened core) to obtain
\be
\Phi_\ell^{\text{in}}(x\to0)\simeq \alpha^{2\ell+\frac{3}{2}}B^{\text{in}}_\ell \left[1+\frac{2\ell(\ell+1)+1}{4(2\ell+3)}\alpha^{2}x^{2}\right]x^{\ell+1}\,,
\label{eq:phinearzero}
\ee
where we corroborate the correction induced to the mode $x^{\ell+1}$ due to the presence of the screened region where the perturbation interacts with the background field. This correction becomes arbitrarily small as we approach the origin, in analogy with the decaying tail in the asymptotic outer region decreasing as we approach $x\to\infty$. This behaviour may motivate to define an inner linear response as
\be
k_{\text{in},\ell}\equiv \frac{2\ell(\ell+1)+1}{4(2\ell+3)}\alpha^{2},
\label{eq:innerpolark}
\ee
that measures the relative weight of the correction. Let us notice that this parameter must be measure sufficiently inside the unscreened score so that the condition $\alpha x\ll1$ for the expansion \eqref{eq:phinearzero} to be valid. As said, we will not explore this regime any further in this work and we will return to our study of the outer asymptotic properties instead. 

\subsection{Ladders and charges}

The equation for the perturbations in the exterior of the sphere is identical to that of a point-like particle. As shown in \cite{BeltranJimenez:2022hvs,BeltranJimenez:2024zmd}, the equations admit a factorization in terms of ladder operators that allow to connect solutions with different $\ell$'s. More explicitly, the hypergeometric form of the perturbation equations outside the sphere given in \eqref{eq:polarhypereq}, can be equivalently written in terms of the Hamiltonian operators
\be
H_{\ell}\equiv -z(1-z)\left[z(1-z)\frac{\dd^2}{\dd z^2}+\frac{3-z}{4}\frac{\dd}{\dd z}+\frac{\ell(\ell+1)-2}{16}\right].
\ee
This family of Hamiltonians admit two ladder structures \cite{BeltranJimenez:2022hvs}:
\begin{itemize}
    \item {\bf{Big ladder}}. The first family of ladder operators is given by
    \bea
\Am_\ell&\equiv& z(z-1)\frac{\dd}{\dd z}-\frac{\ell+6}{4}\left(z-\frac{\ell}{2\ell+5}\right),\\
\Ap_\ell&\equiv&-z(z-1)\frac{\dd}{\dd z}-\frac{\ell-1}{4}\left(z-\frac{\ell+5}{2\ell+5}\right),
\eea
in terms of which we have the factorisation
\bea
\Am_\ell \Ap_\ell=H_\ell+\varepsilon_\ell,\quad\quad
\Ap_\ell \Am_\ell=H_{\ell+4}+\varepsilon_\ell\,,
\label{eq:defApAm}
\eea
with
\be
\varepsilon_\ell=\frac{\ell(\ell+6)(\ell+5)(\ell-1)}{16(2\ell+5)^2}\,.
\ee
This factorisation means that the defined big ladder operators connect solutions of multipoles $\ell$ and $\ell\pm4$. This can be easily shown by noticing the intertwining relations $A^+_\ell H_\ell=H_{\ell+4}A_\ell^+$ and $A_\ell^-H_{\ell+4}=H_\ell A_\ell^-$.

\item {\bf{Small ladder}}. The second family of ladder operators is
\bea
\am_\ell&\equiv& z(z-1)\frac{\dd}{\dd z}+\frac{\ell-5}{4}\left(z-\frac{\ell+1}{2\ell-3}\right)\,,\\
\ap_\ell&\equiv&-z(z-1)\frac{\dd}{\dd z}+\frac{\ell+2}{4}\left(z-\frac{\ell-4}{2\ell-3}\right)\,,
\eea
that satisfy the following relations:
\bea
\am_\ell \ap_\ell=H_\ell+\varepsilon_\ell,\quad\quad \ap_\ell \am_\ell=H_{3-\ell}+\varepsilon_\ell\,,
\eea
with
\be
\varepsilon_\ell=\frac{(\ell-5)(\ell-4)(\ell+2)(\ell+1)}{16(2\ell-3)^2}\,.
\ee
The connection via this small ladder operators is between multipoles $\ell$ and $3-\ell$. Since $\ell$ is non-negative, this means that the small ladder connects the monopole $\ell=0$ to $\ell=3$ and the dipole $\ell=1$ to the quadrupole $\ell=2$. 
\end{itemize}

A detailed analysis of these ladder structures, their properties and some relations to super-symmetric quantum-mechanical systems can be found in \cite{BeltranJimenez:2022hvs}. Here we are only interested in the fact that the small ladder permits to connect the first four multipoles and the big ladder connects these first four multipoles with all the rest. Thus, all multipoles can be connected to the monopole and the dipole.\footnote{There is a subtlety in this connection of the monopole and the dipole with all the rest because of the existence of non-trivial overlap between the kernels of the ladder operators and the Hamiltonian. This non-trivial overlap causes that some ladder operators project out some solutions so there is only a partial connection of solutions. This is discussed at length in \cite{BeltranJimenez:2022hvs,BeltranJimenez:2024zmd} and we will encounter this obstruction also in our analysis below.} We can then use these ladder operators to gain some insights into the linear response coefficients obtained in the previous section. In particular, we will be interested in studying the multipoles that exhibit vanishing linear response coefficients in the point-like limit but acquire a finite size correction. Let us then study how to generate solutions by climbing from the monopole and the dipole.\\

\noindent\bf{Monopole}\\

We start by noticing that the hypergeometric equation \eqref{eq:polarhypereq} that describes the exterior of the sphere for $\ell=0$ can be written as
\be
\frac{\dd}{\dd z}\left[\frac{z^{5/4}}{\sqrt{1-z}}\frac{\dd}{\dd z}\left(\frac{\tilde{\Phi}^{\text{out}}_0}{z^{1/4}}\right)\right]=0,
\ee
which explicitly shows the existence of the following on-shell conserved quantity:
\be
\mQ_{0}^{\text{out}}=\frac{z^{5/4}}{\sqrt{1-z}}\frac{\dd}{\dd z}\left(\frac{\tilde{\Phi}^{\text{out}}_0}{z^{1/4}}\right).
\ee
We can then write the general solution for the monopole as\footnote{Let us notice that the $z$-variable inside and outside the sphere are different. In terms of the globally defined variable $x=r/\rs$, we have $z_{\text{in}}=-\alpha^2 x^2$ while outside $z_{\text{out}}=-x^4$. We will not distinguish between both to alleviate the notation with the understanding that functions and operators in the exterior refer to $z_{\text{out}}$, while interior functions and operators will refer to $z_{\text{in}}$. Whenever a potential confusion arises, we will go back to the global variable $x$.} 
\be
\tilde{\Phi}^{\text{out}}_0=c_0(-z)^{1/4}-4 \mQ^{\text{out}}_0 \;_2F_1\left(-\tfrac12,-\tfrac{1}{4},\tfrac{3}{4}, z\right),
\ee
where $c_0$ is an integration constant. By comparing to the solution given in \eqref{eq:gensolpolout} for $\ell=0$ and using $\;_2F_1\left(-\frac14,0,\frac{5}{4}, z\right)=1$, we can identify $c_0=B_0^{\text{out}}$ and $\mQ_0=-\frac14 A_0^{\text{out}}$ so we can associate the mode $B_0^{\text{out}}$ to the sector with a trivial charge $\mQ^{\text{out}}_0=0$, while the mode $A_0^{\text{out}}$ describes the sector carrying a non-trivial charge $\mQ^{\text{out}}_0\neq0$. Of course, the amount of charge is determined by the behaviour of the solution inside the sphere that determines the boundary conditions for the outer solution. 

The hypergeometric equation \eqref{eq:hyperpolin} that governs the behaviour of the monopole inside the sphere can also be written as the following conservation equation:\footnote{One may wonder if a ladder structure can also be constructed for the interior solutions following the general procedure provided in \cite{BeltranJimenez:2022hvs,BeltranJimenez:2024zmd}. However, an obstruction occurs because the {\it friction} term in \eqref{eq:hyperpolarin} depends on $\ell$ and this feature prevents a factorisation following the method of those references. We will explore this issue in more detail elsewhere.}
\be
\frac{\dd}{\dd z}\left[\sqrt{\frac{z}{z-1}}\frac{\dd \Phi^{\text{in}}_0}{\dd z}\right]=0,
\ee
so we have the interior conserved charge
\be
\mQ_0^{\text{in}}=\sqrt{\frac{z}{z-1}}\frac{\dd \Phi^{\text{in}}_0}{\dd z}.
\ee
We can write the interior solution for the monopole in terms of this conserved charge as
\be
\tilde{\Phi}_0^{\text{in}}=C_0+\mQ_0^{\text{in}} \;_2F_1\left(-\tfrac12,\tfrac{1}{2},\tfrac{3}{2}, -\alpha^2x^2\right)\,.
\ee
We can now apply the ladder operators to generate solutions for other multipoles. We are here interested in the connection of the monopole with the multipoles that exhibit vanishing linear response in the point-like limit. Such multipoles correspond to the tower $\ell=3,7,11,\cdots$ which can be reached by first using the small ladder to obtain $\ell=3$ and then the big ladder to climb up in steps of $\Delta\ell=4$. Then, let us first apply the small ladder to the monopole solution to obtain
\be
\Phit_3^{\text{out}}=a^+_0\Phit_0^{\text{out}}= c_0\frac{3z-5}{12} (-z)^{1/4}+ \frac{8}{3}\mQ^{\text{out}}_0 \;_2F_1\left(-\tfrac54,\tfrac{1}{2},\tfrac{3}{4}, z\right).
\ee
In the point-like limit, regularity at the origin requires $\mQ^{\text{out}}_0=0$ \cite{BeltranJimenez:2022hvs}. This means that the physical solution coincides with the trivially charged sector. Furthermore, the physical mode $c_0$  corresponds to a purely growing solution with no decaying tail at infinity and, thus, the linear response coefficient will vanish. Since the ladder operators connect regular solutions at the origin with regular solutions at the origin and they cannot generate decaying modes from purely growing modes, as evident from their form, we can conclude that all the multipoles connected to $\ell=3$ via $A^+_\ell$ will also have vanishing linear response coefficients in the point-like limit. This is nothing but the observation already made in \cite{BeltranJimenez:2022hvs} in much more detail. For the finite size sphere considered in this work, the boundary condition is imposed at the surface of the sphere by matching to the interior solution. When doing so, we obtain a physical solution in the exterior that has a non-trivial charge $\mQ^{\text{out}}_0$ generated by the interior solution. This non-trivial charge in the exterior will then lead to a non-vanishing linear response that is proportional to $\mQ^{\text{out}}_0$ and this will propagate to all multipoles connected via the ladder.\\

\noindent\bf{Dipole}\\

As in the previous case of the monopole, we first notice that the hypergeometric equation for the dipole outside the sphere can be written as
\be
\frac{\dd}{\dd z}\left[\frac{z^{3/4}}{\sqrt{1-z}}\frac{\dd\tilde{\Phi}^{\text{out}}_{1}}{\dd z}\right]=0,
\ee
from where it immediately follows the existence of the following conserved charge:
\be
\mQ^{\text{out}}_{1}=\frac{z^{3/4}}{\sqrt{1-z}}\frac{\dd \tilde{\Phi}^{\text{out}}_{1}}{\dd z}.
\ee
By integrating this equation we can express the exterior dipole solution as 
\be
\tilde{\Phi}^{\text{out}}_1=c_1+4 \mQ^{\text{out}}_1 z^{1/4} \;_2F_1\left(-\tfrac12,\tfrac{1}{4},\tfrac{5}{4}, z\right).
\label{eq:extdipole}
\ee
Now we can connect the amplitude of the charged and non-charged modes to the interior solution with the appropriate boundary condition. We can notice that the equation for the dipole inside the sphere can also be written as the following conservation law:
\be
\frac{\dd}{\dd z}\left[\frac{z^{5/2}}{\sqrt{1-z}}\frac{\dd}{\dd z}\left(\frac{\tilde{\Phi}^{\text{in}}_{1}}{z^{3/2}}\right)\right]=0,
\ee
so, again, we have a conserved charge, now in the interior, given by
\be
\mQ_1^{\text{in}}=\frac{z^{5/2}}{\sqrt{1-z}}\frac{\dd}{\dd z}\left(\frac{\tilde{\Phi}^{\text{in}}_{1}}{z^{3/2}}\right).
\ee
Integrating this conservation law, we can express the interior dipole perturbation solution as:
\be
\tilde{\Phi}^{\text{in}}_1=C_1(-z)^{3/2}-\frac{2}{3}\mQ_1^{\text{in}}(1-z)^{3/2}.
\ee
This expression shows that the solution with the appropriate boundary condition corresponds to the trivially charged mode with $\mQ_1^{\text{in}}=0$, that is equivalent to the boundary condition $A_1^{\text{in}}=0$. The multipoles with $k_\ell=\mathcal{O}(x_0)$ (i.e., with vanishing linear response in the point-like limit) that are connected to the dipole are those with $\ell=5,9,13,\cdots$ and the connection is directly via the big ladder. An interesting feature of the big ladder is that $A_1^+=-z(z-1)\frac{\dd}{\dd z}$ projects out the mode $c_1$ from the external solution. This was already noticed in \cite{BeltranJimenez:2022hvs} where it was also suggested that a connection between different multipoles can instead be done by descending the ladder rather than climbing up. In the point-like limit, the exterior solution \eqref{eq:extdipole} must be regular at the origin, and this boundary condition selects the charged mode $\mQ_1^{\text{out}}$ and trivialises the mode $c_1$. As explained in \cite{BeltranJimenez:2022hvs}, this property permits to understand the vanishing of linear response coefficients through the connections of the different multipoles via the ladder operators. In our case of a finite sphere, the boundary condition of the external perturbation is obtained by matching at the edge of the sphere with the interior solution. This matching generates a combination of both the trivially charged and the non-trivially charged modes for the outer solution so that the multipoles with vanishing linear response in the point-like case, now acquire a correction due to the finite size of the sphere. 

\section{Axial modes}
\label{Sec:AxialModes}

After studying the polar modes that are common to BI and DBI, we now turn our attention to the axial perturbations, which are genuine of BI and do not exist in DBI. The procedure will be the same as for the polar sector so we will spare some redundant clarifications and discussions for the sake of lightness in the reading.

\subsection{Exterior}

The effective mass for the axial modes in the exterior of the sphere is
\be
\rs^2 m_{\text{A,out}}^2=\frac{\ell(\ell+1)x^2}{1+x^4}+\frac{2+5x^4}{x^2(1+x^2)^2},
\ee
which again coincides with that of a point-like particle. We can recast the equation in the form of a hypergeometric equation by introducing the variable $z=-x^4$ and the redefinition $\tilde{\Psi}_\ell=\tfrac{x}{(1+x^4)^{1/4}}\Psi_\ell$. The resulting equation reads
\be
z(1-z)\frac{\dd^2\tilde{\Psi}^{\text{out}}_\ell}{\dd z^2}+\frac{1-3z}{4}\frac{\dd\tilde{\Psi}^{\text{out}}_\ell}{\dd z}+\frac{\ell(\ell+1)}{16}\tilde\Psi^{\text{out}}_\ell=0,
\label{eq:axialhypereqout}
\ee
which can be solved in term of hypergeometric functions that yield the following solution for the original variable 
\be
\Psi^{\text{out}}_\ell=\left(1+x^4\right)^{1/4}\left[\frac{A^{\text{out}}_{\ell}}{x}  \;_2F_1\left(-\tfrac{\ell+1}{4},\tfrac{\ell}{4},\tfrac{1}{4}, -x^4\right)+B^{\text{out}}_{\ell}\;x^2  \;_2F_1\left(\tfrac{2-\ell}{4},\tfrac{\ell+3}{4},\tfrac{7}{4},-x^4\right)\right]\,,
\label{eq:solPsigenout}
\ee
with $A^{\text{out}}_\ell$ and $B^{\text{out}}_\ell$ the integration constants to be set by boundary conditions.

\subsection{Interior}

The effective mass in the interior of the sphere is given by
\be
\rs^2 m_{\text{A,in}}^2=\frac{\ell(\ell+1)}{(1+\alpha^2x^2)x^2}+\frac{(2-\alpha^2x^2)\alpha^2}{4(1+\alpha^2 x^2)^2}.
\ee
Once again, we can introduce a radial variable $z=-\alpha^2 x^2$ and a field redefinition $\tilde{\Psi}_\ell=\tfrac{x^\ell}{(1+\alpha^2x^2)^{1/4}}\Psi_\ell$ to recast the multipole equations as the following hypergeometric equation
\be
z(1-z)\frac{\dd^2\tilde{\Psi}^{\text{in}}_\ell}{\dd z^2}+\frac{2(\ell-1)z+1-2\ell}{2}\,\frac{\dd\tilde{\Psi}^{\text{in}}_\ell}{\dd z}-\frac{\ell^2}{4}\tilde\Psi^{\text{in}}_\ell=0.
\label{eq:axialhypereqin}
\ee
The solution can then be written as 
\begin{align}
\Psi^{\text{in}}_\ell=\frac{\left(1+\alpha^2x^2\right)^{1/4}}{\sqrt{\alpha}}
\bigg[&A^{\text{in}}_{\ell}x^{-\ell} \;_2F_1\left(-\tfrac{\ell}{2},-\tfrac{\ell}{2},\tfrac{1-2\ell}{2}, -\alpha^2x^2\right)\nonumber\\
+&\,B^{\text{in}}_{\ell}\;x^{\ell+1}  \;_2F_1\left(\tfrac{\ell+1}{2},\tfrac{\ell+1}{2},\tfrac{2\ell+3}{2},-\alpha^2x^2\right)\bigg]\,.
\label{eq:solPsigenin}
\end{align}
To fix the integration constants $A^{\text{in}}_{\ell}$ and $B^{\text{in}}_{\ell}$ we will impose the vanishing of the potential at the origin and this amounts to requiring $A^{\text{in}}_{\ell}=0$. Thus, the interior solution with the appropriate boundary condition reduces to
\be
\Psi^{\text{in}}_\ell=\alpha^{\ell-\tfrac{1}{2}}B^{\text{in}}_{\ell}\left(1+\alpha^2x^2\right)^{1/4}
\;(\alpha x)^{\ell+1}  \;_2F_1\left(\tfrac{\ell+1}{2},\tfrac{\ell+1}{2},\tfrac{2\ell+3}{2},-\alpha^2x^2\right)\,.
\label{eq:solPsigeninreg}
\ee
Now that we have the solution in both regions, we can proceed to computing the linear response coefficients.

\subsection{Axial linear response}

\begin{figure}[ht]
\includegraphics[width=0.49\linewidth]{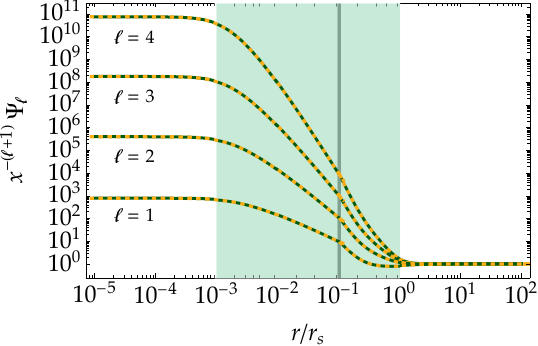}
\includegraphics[width=0.49\linewidth]{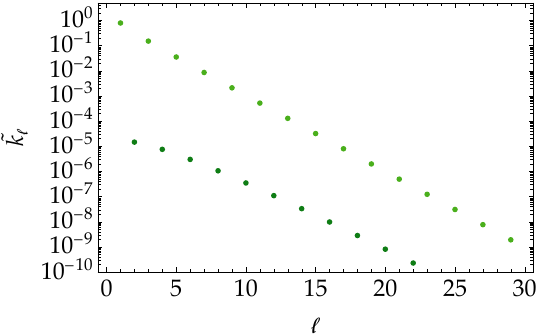}
\caption{Same as in Fig. \eqref{Fig:Axial} for the axial perturbations. In the right panel we can see again a hierarchy between even and odd multipoles. As explained in the main text, this hierarchy is more pronounced for the axial perturbations since the correction due to finite size effects scales as $x_0^3$ instead as the correction $x_0$ obtained for the polar case.
}
\label{Fig:Axial}
\end{figure}

As for the polar sector, the asymptotic solution of the perturbations will feature a growing and a decaying modes corresponding to the external perturbation and the response of the sphere. For the axial perturbations we find

\begin{align}
\Psi_\ell^{\text{out}}(x)\simeq & \,\Gamma\left(-\tfrac{2\ell+1}{4}\right)\left[ \frac{\Gamma(\tfrac{1}{4})}{\Gamma(-\tfrac{\ell+1}{4})\Gamma(\tfrac{1-\ell}{4})}A^{\text{out}}_\ell+\frac{\Gamma(\tfrac{7}{4})}{\Gamma(\tfrac{2-\ell}{4})\Gamma(\tfrac{4-\ell}{4})}B^{\text{out}}_\ell\right]x^{-\ell}\\ &+\Gamma(\tfrac{2\ell+1}{4})\left[\frac{\Gamma(\tfrac{1}{4})}{\Gamma(\tfrac{\ell}{4})\Gamma(\tfrac{\ell+2}{4})}A^{\text{out}}_\ell+\frac{\Gamma(\tfrac{7}{4})}{\Gamma(\tfrac{\ell+3}{4})\Gamma(\tfrac{\ell+5}{4})}B^{\text{out}}_\ell\right]x^{\ell+1},
\end{align}
so we can define the linear response as
\be
\kt\equiv\frac{\Gamma\left(-\tfrac{2\ell+1}{4}\right)\left[ \frac{\Gamma\left(\tfrac{1}{4}\right)}{\Gamma\left(-\tfrac{\ell+1}{4}\right)\Gamma\left(\tfrac{1-\ell}{4}\right)}A^{\text{out}}_\ell+\frac{\Gamma\left(\tfrac{7}{4}\right)}{\Gamma\left(\tfrac{2-\ell}{4}\right)\Gamma\left(\tfrac{4-\ell}{4}\right)}B^{\text{out}}_\ell\right]}{\Gamma\left(\tfrac{2\ell+1}{4}\right)\left[\frac{\Gamma\left(\tfrac{1}{4}\right)}{\Gamma\left(\tfrac{\ell}{4}\right)\Gamma\left(\tfrac{\ell+2}{4}\right)}A^{\text{out}}_\ell+\frac{\Gamma\left(\tfrac{7}{4}\right)}{\Gamma\left(\tfrac{\ell+3}{4}\right)\Gamma\left(\tfrac{\ell+5}{4}\right)}B^{\text{out}}_\ell\right]}.
\ee
In order to evaluate this expression, we need to match to the interior and exterior solutions at the surface of the sphere. As in the polar case, the effective mass contains a Dirac delta function localised at the surface of the sphere that generates a discontinuity in the derivative of $\Psi_\ell$ at $x=x_0$. Proceeding analogously to the polar sector we obtain
\be
(\Delta\Psi_\ell')_{x_0}=\lim_{\epsilon\to0}\int_{x_0-\epsilon}^{x_0+\epsilon}\mA^2(x)\Psi_\ell(x)\dd x=-\frac{3}{2x_0}\Psi_\ell(x_0).
\ee
Thus, the appropriate matching conditions are given by
\be
\Psi_\ell^{\text{in}}(x_0)=\Psi_\ell^{\text{out}}(x_0),\qquad \left[\frac{\dd\Psi_\ell^{\text{in}}}{\dd x}-\frac{\dd\Psi_\ell^{\text{out}}}{\dd x}\right]_{x_0}=-\frac{3}{2x_0}\Psi_\ell(x_0)\,,
\ee
which correctly reproduce the numerical results shown in Fig. \ref{Fig:Axial}. After imposing these conditions, there remains one integration constant that is fixed by the amplitude of the external perturbation. This remaining constant is again irrelevant for the linear response because it factors out from the exterior solution so it does not contribute to linear response coefficient. The full expression for $\kt_\ell$ is again very contrived so we will simply show their numerical values in Fig.~\ref{Fig:Axial}. The expansion for a small sphere as compared to its screening radius, i.e. $x_0\ll1$, leads to 
\be
\kt_\ell\simeq\kt_\ell^{(0)}+\kt_\ell^{(3)}x_0^3
\ee
with 
\begin{eqnarray}
    \kt_\ell^{(0)}&=&\frac{\Gamma\left(-\frac{2\ell+1}{4}\right)\Gamma\left(\frac{\ell+3}{4}\right)\Gamma\left(\frac{\ell+5}{4}\right)}
{\Gamma\left(-\frac{\ell-2}{4}\right)\Gamma\left(-\frac{\ell-4}{4}\right)\Gamma\left(\frac{2\ell+1}{4}\right)},\\
\kt_\ell^{(3)}&=&\frac{2^{-\ell}}{\pi}\frac{\Gamma\left(\tfrac{5}{4}\right)\Gamma\left(-\tfrac{2\ell+1}{4}\right)\Gamma^2\left(\tfrac{\ell+3}{2}\right)}{\Gamma\left(\tfrac{7}{4}\right)\Gamma\left(\tfrac{2\ell+1}{4}\right)}\left[\sin\left(\frac{\ell\pi}{2}\right)+\cos\left(\frac{\ell\pi}{2}\right)\right]\beta(x_0)
\end{eqnarray}
where we have defined the logarithmic running $\beta$ as
\be
\beta( x_0)=1+3\sum_{k=1}^\ell\frac{1}{k}+\log\frac{x_0^6}{8}\,.
\ee
The zeroth-order contribution coincides with the linear response of the point-like particle obtained in \cite{BeltranJimenez:2022hvs}, where it was discussed how the even multipoles above the monopole have vanishing linear response, i.e., $\kt^{(0)}_{2n}=0$ for $n=1,2,3,\dots$. For those multipoles, the leading order contribution is provided by $\kt_\ell^{(3)}$. Interestingly, this leading order scales as $x_0^3$, with a mild logarithmic correction, so it is substantially suppressed with respect to the linear response of the polar modes, whose next-to-leading order (which is in turn the leading order for odd multipoles above the dipole) scales as $x_0$. Let us notice that, although the logarithm in the running function $\beta(x_0)$ diverges for $x_0\to0$, the linear response does not because the $x_0^3$ goes to zero more rapidly, i.e., $x_0^3\log x_0^6$ vanishes as $x_0\to0$. In that limit, the linear response is given by $\tilde{k}_\ell^{(0)}$, thus reproducing the result of the point-like particle. Again, the fact that the linear response produces a smooth $x_0\to0$ limit is related to the non-singular nature of the BI background field.

\subsection{Axial inner response}

As for the polar sector, the solution for the axial perturbations in the unscreened score will be deformed by the presence of the external stimuli due to the presence of the screened shell. We are then motivated to define an axial inner response in analogy to \eqref{eq:innerpolark}. For the linear theories, the solution has the same structure as for the polar sector, i.e., 
\be
\Psi_\ell^{\text{lin}}=A^{\text{lin}}_\ell x^{-\ell}+B^{\text{lin}}_\ell x^{\ell+1},
\ee
where the mode $A^{\text{lin}}_\ell$ is trivialised by imposing regularity at the origin. The idea is then to compare this solution for the linear theory with the perturbation of the non-linear theories in the unscreened core. The regular solution of (D)BI near the origin has the following the expansion:
\be
\Psi_\ell^{\text{in}}(x\to0)\simeq \alpha^{2\ell+\frac{1}{2}}B^{\text{in}}_\ell \left[1+\frac{1-2\ell(\ell+1)}{4(2\ell+3)}\alpha^2x^2\right]x^{\ell+1}\,,
\label{eq:psinearzero}
\ee
that shows the correction to the mode $x^{\ell+1}$ due to the screened shell where the non-linearities are important. Thus, we are motivated to the define the axial linear response as
\be
\tilde{k}_{\text{in},\ell}\equiv \frac{1-2\ell(\ell+1)}{4(2\ell+3)}\alpha^2\,,
\ee
which, as in the polar case, is to be measured well inside the unscreened core where $\alpha x\ll1$. Unlike the deformation in the outer region, where the linear response coefficients of both sectors have a different structure, the coefficients defined from the deformation in the unscreened core share the same structure for both sectors. Again, we will not explore this regime any further here and will proceed to discussing the ladder structure of the axial sector.

\subsection{Ladders and charges}
The axial sector of BI perturbations around a point-like particle admits an analogous 2-ladder structure to the polar sector \cite{BeltranJimenez:2022hvs} and this property will also be shared by the exterior solutions of the finite size sphere. The starting point is again expressing the perturbation equations in terms of the Hamiltonians
\be
H_\ell\equiv -z(1-z)\left[z(1-z)\frac{\dd^2}{\dd z^2}+\frac{1-3z}{4}\frac{\dd}{\dd z}+\frac{\ell(\ell+1)}{16}\right],
\ee
and factorize it in terms of ladder operators. As in the polar case, this family of Hamiltonians admits two ladder structures:

\begin{itemize}
    \item \bf{Big ladder}. The operators corresponding to the big ladder read
\bea
\Bm_\ell&\equiv& z(z-1)\frac{\dd}{\dd z}-\frac{\ell+5}{4}\left(z-\frac{\ell+3}{2\ell+5}\right),\\
\Bp_\ell&\equiv&-z(z-1)\frac{\dd}{\dd z}-\frac{\ell}{4}\left(z-\frac{\ell+2}{2\ell+5}\right),
\eea
which give rise to the following factorisation:
\bea
\Bm_\ell \Bp_\ell=H_\ell+\delta_\ell,\quad\quad \Bp_\ell \Bm_\ell=H_{\ell+4}+\delta_\ell,
\eea
with
\be
\delta_\ell=\frac{(\ell+5)(\ell+3)(\ell+2)\ell}{16(2\ell+5)^2}.
\ee
As in the polar case, the big ladder connects solutions for the perturbation of multipole $\ell$ to solutions of multipole $\ell\pm4$.

\item \bf{Small ladder.} The other ladder structure is provided by the operators
\bea
\bm_\ell&\equiv& z(z-1)\frac{\dd}{\dd z}-\frac{\ell-4}{4}\left(z+\frac{\ell-2}{3-2\ell}\right),\\
\bp_\ell&\equiv&-z(z-1)\frac{\dd}{\dd z}-\frac{\ell+1}{4}\left(z+\frac{\ell-1}{3-2\ell}\right),
\eea
that produce the factorisation
\bea
\bm_\ell \bp_\ell=H_\ell+\delta_\ell,\quad\quad \bp_\ell \bm_\ell=H_{3-\ell}+\delta_\ell,
\eea
with
\be
\delta_\ell=\frac{(\ell-4)(\ell-2)(\ell-1)(\ell+1)}{16(2\ell-3)^2}.
\ee
Again, this small ladder represents a connection among the first four multipoles.
\end{itemize}
Thus, we find again that the monopole and the dipole can be connected with the rest of multipoles by climbing up with both ladder operators. The story is then relatively similar to the polar case. The exterior solutions for the monopole and the dipole can be classified in terms of the conserved charges obtained in \cite{BeltranJimenez:2022hvs} that also allow to relate the properties of the linear response coefficients with the charges and the relations of the multipoles via the ladder structures. Again, the interior solutions for the monopole and the dipole can also be expressed in terms of conserved charges that permit to classify into physical and non-physical solutions. These charges in the interior affect the exterior solutions through the matching at $x=x_0$ and this gives corrections to $\tilde{k}_\ell$ with respect to the point-like that can be connected via the ladder operators.

\section{Discussion}
\label{Sec:Discussion}

The goal of this work has been to extend previous studies of a point-like particle within BI electromagnetism and DBI scalar theories to finite-size spherically symmetric objects. In particular, our main interest has been to analyse how the vanishing of certain (static) linear response coefficients for the point-like particle is affected when considering finite size corrections. To compute the linear response we have first obtained the background field profile generated by the sphere outside and inside. On top of this background field we have computed the equations for the perturbations that we have decomposed into multipoles to exploit the background spherical symmetry. In order to solve the equations, we have considered the inner and outer regions, where the equations admit analytic solutions in terms of hypergeometric functions, and we have matched at the surface of the sphere. The matching conditions required imposing continuity of the solutions but a discontinuity in their first derivatives Regarding boundary conditions, we have imposed regularity at the center of the sphere and matching the amplitude of the external perturbation at infinity. We have also solved the equations numerically and found perfect agreement with our analytical solutions. After obtaining the solutions for the perturbations we have been able to compute the linear response coefficients.

The main result obtained from our study is that the multipoles with vanishing linear response in the point-like case cease to be irresponsive and acquire a correction due to the finite size of the object. This result shows that the vanishing of the linear response coefficients is tightly linked to the point-like character. Interestingly, although the point-like particle limit can be subtle to take in intermediate steps, we have found that the linear response coefficients in turn have a smooth vanishing size limit. Although this was not guaranteed a priory, the remarkable property of BI and DBI of having a non-singular field for a point-like particle at the position of the particle permits to have a smooth limit. We have computed the finite size effects for polar perturbations (that are common to BI and DBI) and for axial perturbations (which only exist for BI). For the polar modes we have obtained that the leading order correction to the response coefficients goes as the radius of the sphere, while the correction for the axial modes goes as the volume of the sphere (with some mild logarithmic correction). This result suggests that the axial perturbations are less sensitive to finite size effects than the polar sector. The fact that some multipoles have linear response coefficients with a leading order governed by the screening scale, while other multipoles have a leading order determined by the radius of the sphere introduces a hierarchy that can be exploited to measure the properties of the object at infinity. In the same phenomenological vein and for dark matter models based on non-linear electromagnetism, we have also pointed out that the absence of screening deep inside the sphere and the response coefficients in this inner region may be traced by studying the behaviour of galaxies well inside a cluster.

%%%%%%%%%%%%%%%%%%%%%%%%%%%%
\section*{Acknowledgements}

JBJ and DB acknowledge support from grants PID2021-122938NB-I00 and PID2024-158938NB-I00 funded by MICIU/AEI/10.13039/501100011033 and by “ERDF A way of making Europe”, the Project SA097P24 funded by Junta de Castilla y Le\'on and the research visit grant PRX23/00530. JBJ thanks the Institute of Theoretical Astrophysics at the University of Oslo for their hospitality during the completion of this work.
%%%%%%%%%%%%%%%%%%%%%%%%%%%%

%%%%%%%%%%%%%%%%%%%%%%%%%%%%%%%%%
\bibliography{bibFinitesize}
%%%%%%%%%%%%%%%%%%%%%%%%%%%%%%%%%

\end{document}